# Zero-Dimensional Stacking Domains Enable Strong-Ductile Synergy in Additive Manufactured Titanium

**Authors:** Wenjing Zhang[1+], Jizhe Cui[1,2,3+], Xiaoyang Wang[4*], Shubo Zhang[1], Yan Chong[5, 6*], Andy Godfrey[1], Nobuhiro Tsuji[5], Kai Wang[7], Rong Hu[8, 9], Jing Xue[9], Junyu Chen[10], Gang Fang[10], Rong Yu[1,2,3*], Wei Liu[1,6*]

**Affiliations:**

[1]*School of Materials Science and Engineering, Tsinghua University, Beijing 100084, China.*
[2]*Key Laboratory of Advanced Materials of Ministry of Education, Tsinghua University, Beijing 100084, China.*
[3]*State Key Laboratory of New Ceramics and Fine Processing, Tsinghua University, Beijing 100084, China.*
[4]*Laboratory of Computational Physics, Institute of Applied Physics and Computational Mathematics, Beijing 100094, China*
[5]*Department of Materials Science and Engineering, Kyoto University, Kyoto 606-8501, Japan.*
[6]*Structural Materials Division, Suzhou Laboratory, Jiangsu 215123, China.*
[7]*School of Materials Science and Engineering, Beihang University, Beijing 100191, China*
[8]*Key Laboratory for Light-weight Materials, Nanjing Tech University, Nanjing 210009, PR China*
[9]*Analysis and Characterization Platform, Advanced Materials Research Institute, Yangtze Delta, Suzhou 215133, China*
[10]*State Key Laboratory of Clean and Efficient Turbomachinery Power Equipment, Department of Mechanical Engineering, Tsinghua University, Beijing 100084, China*

+ These authors contributed equally to this work

Corresponding authors: wang_xiaoyang@iapcm.ac.cn, chongy@szlab.ac.cn(Yan Chong), ryu@tsinghua.edu.cn (Rong Yu), liuw@tsinghua.edu.cn(Wei Liu)

## Abstract

Alloying by addition of oxygen interstitials during additive manufacturing provides new routes to strengthen and toughen metals and alloys. The underlying mechanisms by which such interstitial atoms lead to enhanced properties remain, however, unclear, not least due a lack of quantitative atomic-scale models linking microstructure to properties. Here using quasi-3D imaging based on multi-slice electron ptychography, we reveal the importance of a new type of interstitial-character lattice defect, namely zero-dimensional stacking domains (ZDSDs), present in high density in AM-processed oxygen-modulated pure titanium. These ZDSDs promote slip diversity, and support intense work hardening, enabling a three-fold enhancement in both strength and ductility in Ti-0.45O compared to conventional pure Ti. The work demonstrates the potential for using interstitial solutes to enhance mechanical properties in a range of critical engineering alloys.

Titanium has emerged as a critical material for engineering structures and medical implants due to its low density, exceptional corrosion resistance, and superior biocompatibility (1-6). Alloying elements such as aluminum (Al), chromium (Cr), molybdenum (Mo), vanadium (V), and niobium (Nb) are typically added to enhance strength (5). However, such alloying content increases cost and impedes recyclability, conflicting with sustainability principles (7). Moreover, although elements like Al and V improve strength, they pose neurotoxicity risks linked to Alzheimer's disease and osteomalacia (8, 9). Consequently, there is a pressing need to develop novel titanium-based materials that simultaneously achieve high strength, ductility, biosafety, and cost-effectiveness, thereby enabling use of such alloys in high end applications.

As an alternative, alloying of titanium with light interstitial elements such as oxygen, nitrogen and carbon, offers a promising pathway toward enhanced mechanical properties (*1, 3, 5*). Interstitial hardening is notably more potent and cost-effective than substitutional solute hardening (e.g., via Al, V, or Cr). For example, adding just 1.0 at.% oxygen elevates the yield strength of titanium to a level comparable to that from addition of 10 at.% Al (*10*). However, interstitial alloying is hindered by severe embrittlement for oxygen content exceeding 0.3 wt.%. The underlying mechanism for this oxygen-induced embrittlement lies in localized plastic deformation caused by planar slip (*1*) and grain boundary susceptibility to oxygen segregation (*4, 11*).

Additive manufacturing (AM), leveraging ultrahigh cooling rates ($10^6$–$10^7$ K/s) and non-equilibrium solidification, provides a novel pathway to regulate oxygen distribution and mitigate high-oxygen embrittlement in pure titanium. Many studies have now confirmed that AM-processed oxygen-modulated titanium can achieve simultaneous strength-ductility enhancements (*8, 9, 11-18*). Proposed mechanisms for this include grain-size effects (grain refinement induced plasticity(*13, 18*), and grain boundary serrations promoting multi-slip activity while suppressing intergranular cracking) (*9*), and second-phase effects. Regarding the latter, Ding et al. (*17*) attributed the strength-ductility synergy to the introduction of FCC phase regions within the HCP matrix, with concomitant oxygen redistribution between the two phases. Crucially, all existing strengthening-toughening hypotheses are based on known microstructural features (dislocations, grain boundaries, second phases) and lack atomically quantitative structure-property models. This gap fundamentally obscures the origin of the strength-ductility synergy in AM oxygen-modulated titanium.

In this work, using super-resolution adaptive propagation ptychography (APP), we reveal the presence of previously unreported high-density zero-dimensional stacking domains (ZDSDs) in oxygen-modulated pure titanium fabricated by laser powder bed fusion (LPBF) – the predominant metal AM technique. These nanoscale configurations feature self-interstitial structures coherent with the Ti matrix. Multi-scale characterization and computational simulations reveal the formation mechanism of the ZDSDs, together with their interaction with oxygen interstitials and dislocations, and their critical role in simultaneous strengthening and toughening. By optimizing oxygen content and processing parameters, we achieve an ultimate tensile strength of 915 MPa with 15.6% uniform elongation in a Ti-0.45O alloy – a near threefold enhancement in strength-ductility synergy over conventional pure titanium. This discovery challenges fundamental assumptions in physical metallurgy and establishes a paradigm-shift in structure-property relationships.

**Zero-dimensional stacking domain (ZDSD) in LPBF Ti-O alloys**

Several LPBF Ti-O alloys with different oxygen content, namely Ti-0.09O, Ti-0.13O, Ti-0.24O, Ti-0.33O and Ti-0.45O (wt.%) were prepared by laser powder bed fusion (LPBF) (*19*). The as-printed Ti-O alloys (0.09-0.45 wt.% O) exhibit a single-phase α (HCP) microstructure with near-constant grain size (8.8-13.2 μm), lower initial dislocation density at higher oxygen content, and homogeneous oxygen distribution without oxide particles (**Supplementary Text 1** and **figs. S1** to **S4**).

A typical scanning transmission electron microscopy high-angle annular dark-field (STEM-HAADF) image and an annular bright-field (ABF) image of the LPBF Ti-0.45O alloy (using a [11-20] zone axis) is shown in **Fig. 1, A and B)** (Technique-specific characteristics and the comparative advantages of each method are documented in the **Supplementary Text 2**). The HAADF image exhibits a highly uniform Z-contrast, whereas the ABF image reveals a certain degree of microstructural inhomogeneity and the existence of a high-density of nanostructured features (hereafter nanostructures). A Fourier transform (insert of **Fig. 1B**) clearly reveals extra-spots at (0001) (indicated by red circles in the insert of **Fig. 1B**). An inverse Fourier transform (**fig. S5A)** using these extra spots indicates that they originate from the nanostructures. The average size of the nanostructures is ~2.1 nm, with a volume fraction of 11.5% (**Fig. 1G**) (*19*). In comparison, STEM-HAADF/ABF imaging reveals a low density of nanostructures in Ti-0.09O (**Fig. 1, D** and **E**, **fig. S5B**). Specifically, Fourier transforms of ABF images show markedly weaker (0001) signal intensity versus Ti-0.45O (**Fig. 1E** inset), quantified in intensity profiles (**Fig. 1F**). The ~1.2% nanostructure volume fraction in Ti-0.09O (**Fig. 1G**) (*19*), is far below that in Ti-0.45O, highlighting the important role played by oxygen solutes in promoting the formation of a high-density of nanostructures in LPBF Ti-O alloys.

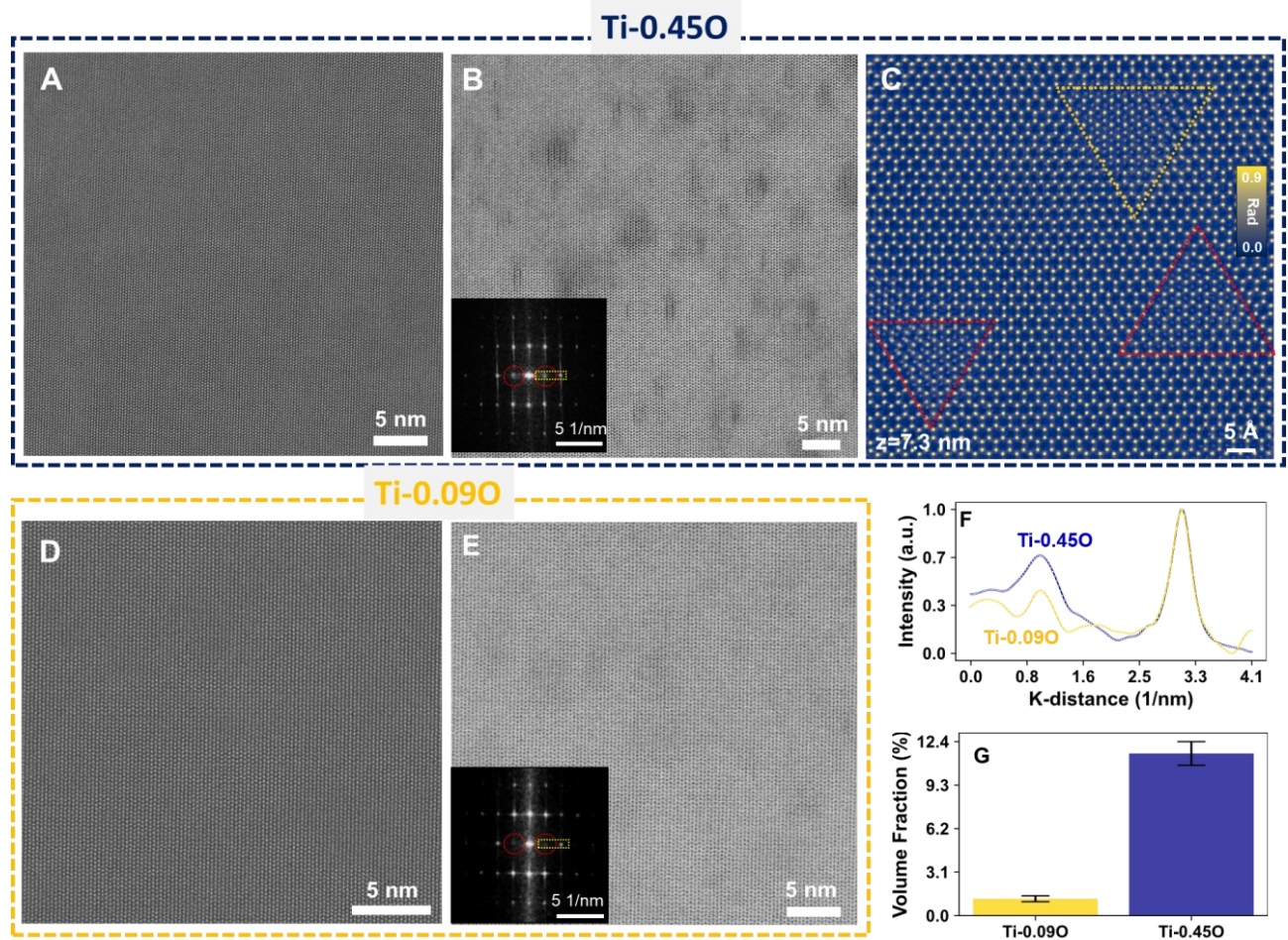

**Fig. 1. Nanostructures (ZDSDs) in LPBF Ti-O alloys**. (**A**) STEM-HAADF image and (**B**) annular bright-field (ABF) image of typical microstructure of LPBF Ti-0.45O. The zone axis is [11-20]. Inset is the fast Fourier transformation (FFT) pattern, in which extra spots at ½ (0002) are indicated by red circles. (**C**) Ptychographic image (with a Z depth of 7.3 nm) along the [0001] zone axis, in which three ZDSDs (highlighted by dashed triangles) are observed in the area of view. (**D**) STEM-HAADF image and (**E**) STEM-ABF image of LPBF Ti-0.09O viewed along [11-20]. Inset is the FFT pattern. (**F**) Diffraction intensities of the rectangular regions in the FFT patterns of Ti-0.45O (blue lines) and Ti-0.09O (yellow line). (**G**) Volume fractions of ZDSDs in Ti-0.45O and Ti-0.09O calculated over 10 different regions.

In the following, we demonstrate that these nanostructures represent a previously unreported class of lattice defect, namely zero-dimensional stacking domains (ZDSDs), distinct from conventional two-dimensional stacking faults and nanotwins. Their atomic structure, formation mechanism, and mechanical properties effect in Ti-O alloys will be discussed.

The 4D-STEM-based adaptive propagator ptychography (APP) technique provides ultra-high spatial/phase resolution and depth-resolution capability to mitigate the interference from surface layer damage and probe intrinsic structural information within the material (**Supplementary Text 2** and **Figs. S6** and **S7**) (*20-23*). A typical APP image of LPBF Ti-0.45O is shown in **Fig. 1C**, in which three ZDSDs (indicated by dashed triangles) are observed in the field of view. Reconstructed images of ZDSDs along [0001] and [11-20] zone axes are shown in **Fig. 2, A** and **B**, respectively. Along the [0001] zone axis (**Fig. 2A**), where HCP-Ti exhibits an *ABABAB…* stacking sequence, additional atomic columns are present at the center of hexagonal rings formed by the atoms in *A* and *B* layers (see dashed yellow-triangle), while the contrast of the *B*-layer atoms is diminished. Moreover, along the [11-20] zone axis, some *B* atomic columns are observed to split into two slightly weaker columns (see dashed rectangular region in **Fig. 2B**). Depth profiles along the green dashed lines in **Fig. 2, A** and **B** are shown in **Fig. 2, C** and **D,** respectively. It is clearly seen that the ZDSDs exhibit a triangular prism structure, with triangle length of ~2 nm and a thickness of ~1 nm. This structural characteristic reflects the shift of *B-layer* atoms to *C* sites within a localized triangular prism, forming a *BC*-stacking domain within the *AB*-stacking matrix of HCP-Ti (**Fig. 2G**). Corresponding schematic illustrations of the atom positions viewed along the [0001] and [11-20] zone axes are shown in **Fig. 2, E** and **F**, respectively. It should be noted that the interface between each ZDSD and surrounding matrix is fully coherent.

The displacement of *B*-layer atoms to *C*-layer sites results in interatomic distances between some *B*- and *C*-layer atoms being shorter than those in defect-free regions. As such, two possible atomic coordination mechanisms may occur at the ZDSD boundary. The first is that excessively close *B*- and *C*-layer atoms are eliminated, resulting in a locally reduced atomic density at the boundary compared to bulk regions. In this case, ZDSDs will have the characteristics of a vacancy-type cluster (**fig. S8B**). The second possibility is that the close distance of *B*- and *C*-layer atoms is permitted, leading to a locally increased atomic density relative to the matrix, implying that the ZDSDs instead manifest as a self-interstitial ZDSDs (**fig. S8C**).

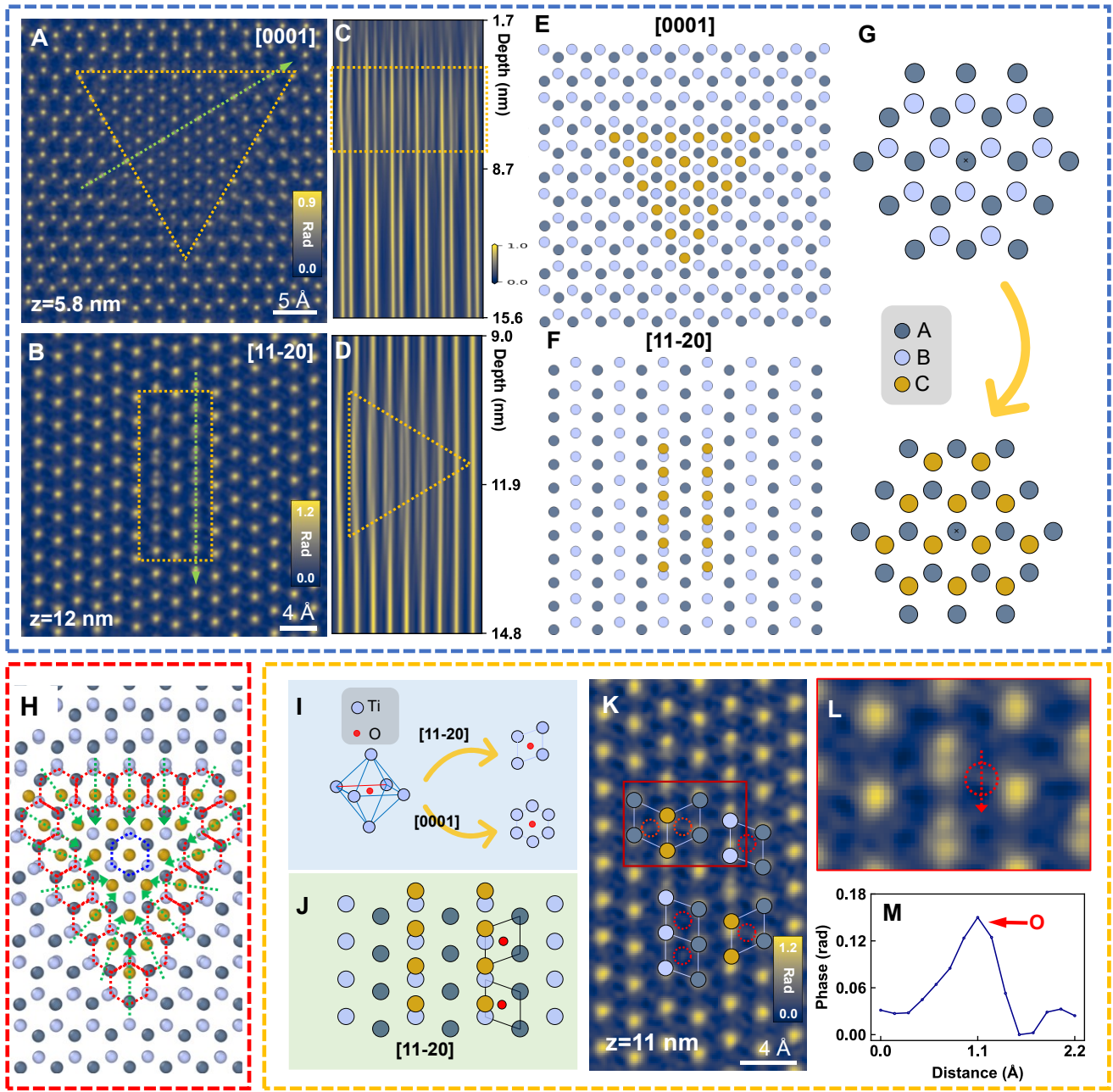

**Fig. 2. Characteristics of ZDSDs in LPBF Ti-0.45O revealed by APP**. Ptychographic images from [0001] (**A**) and [11-20] (**B**) zone axes, in which ZDSD regions are highlighted by yellow dashed lines. The depth for (**A**) and (**B**) correspond to distances from the top surface of the sample of 5.8 nm and 12 nm, respectively. Depth profiles along the green dashed lines in (**A**) and (**B**) are shown in (**C**) and (**D**), respectively, revealing a triangular prism structure of the ZDSDs in 3-dimensions. Schematic illustrations of the ZDSD structure, viewed along [0001] and [11-20], are shown in (**E**) and (**F**), respectively, corresponding well with the experimental observations in (**A**) and (**B**). (**G**) Schematic illustration of the formation of a ZDSD. (**H**) Atomic structure of a self-interstitial-type nano-twin structure along the [0001] zone axis as obtained from molecular dynamics simulations, showing excellent agreement with the APP results in (**A**). (**I–M**) Direct observation of interstitial

oxygen near a ZDSD boundary. **(I)** Schematic illustration of octahedral site occupancy of interstitial oxygen in the HCP lattice, as projected along [0001] and [11-20]. **(J)** Schematic showing possible interstitial oxygen positions in the atomic model containing a ZDSD, viewed along [11-20]. **(K)** APP phase image from a depth of 11 nm within a ZDSD (same area as in panel **(B)**). **(L)** Magnified view of the region enclosed by the red rectangle in **(K)**. **(M)** Phase intensity profile along the red arrow in **(L)**, clearly revealing a distinct signal peak corresponding to interstitial oxygen.

To determine the intrinsic nature of ZDSDs, we employed atomic-scale simulations using two different modeling approaches. Due to high atomic density in boundary regions of self-interstitial ZDSDs, $B$-layer and $C$-layer atoms experience compression. Consequently, the $C$-layer atoms at the ZDSD boundary are pushed toward inward rather than occupying the hexagonal center formed by $A$ and $B$-layer atoms (green arrow in **Fig. 2H**). Such atomic-scale shifts exhibit excellent agreement with the experimental observations in **Fig. 2A**, whereas a vacancy-type ZDSD model fails to reproduce the observationsc (**fig. S8**), indicating that ZDSDs are self-interstitial clusters.

In most metal additive manufacturing processes, the higher density of the solid phase compared to the liquid phase usually leads to the formation of vacancy-type defects or nanovoids during rapid cooling. However, our additive manufactured Ti contains a significant density of interstitial-type ZDSDs, this difference can be attributed to the high-temperature β phase (BCC) of Ti at ambient pressure. Our previous experimental results show that the β phase has a higher density than the α phase (*24*). Therefore, interstitial ZDSDs may form during rapid cooling as the β phase transforms to the α phase. The accompanying lattice expansion cannot be fully accommodated by atomic diffusion on such short timescales, causing excess atoms to remain in the matrix as self-interstitials, which are subsequently retained as interstitial-type ZDSDs.

The **fig. S9** shows ZDSDs exhibit lower formation energies than conventional interstitial defect clusters — including <a>-type (irradiation-induced self-interstitial atom (SIA)) and <c+a>-type (ZDSD-derived in the present study) prismatic loops — for clusters below 18 atoms. However, beyond a specific atom threshold, ZDSDs become energetically unfavorable due to quadratic stacking-fault area scaling versus linear scaling in prismatic loops. **This energetic preference** reveals the formation of small-sized interstitial-type ZDSDs, and lack of observations of very large-sized ZDSDs in our experimental result (**Supplementary text 3**).

The APP results indicate that the formation of ZDSDs is closely related to the addition of oxygen during the fabrication process. **Fig. 2I** illustrates the projected positions of O atoms occupying octahedral interstitial sites, as viewed along [0001] and [11-20]. Due to the overlap between *C*-layer atoms and interstitial oxygen in the [0001] projection, making them indistinguishable, we chose to analyze the structure along the [11-20] zone axis. **Fig. 2K** shows a phase image at a depth of 11 nm within the ZDSD region from **Fig. 2B**, highlighting typical interstitial oxygen signals. These observations align well with the proposed interstitial oxygen positions near the ZDSD, as illustrated in **Fig. 2J**. The magnified view in **Fig. 2L** and the intensity profile along the red arrow in **Fig. 2M** further confirm the presence of interstitial oxygen. Additionally, our first-principles calculations further reveal strong binding (140 meV/oxygen interstitial) between oxygen interstitials and ZDSDs (*19*), indicating oxygen interstitials can effectively reduce the formation energy of ZDSDs, thereby promoting their formation. This result agrees well with the experimental tendency that increased oxygen concentration leads to a higher density of ZDSDs.

**Mechanical properties of LPBF Ti-O alloys**

Representative engineering stress-strain curves of LPBF Ti-O alloys with different oxygen content at room temperature are shown in **Fig. 3A**, together with the average yield strength (YS, $\sigma_y$) and ultimate tensile strength (UTS, $\sigma_u$) measured over at least three specimens for each alloy (**fig. S10**). The corresponding true stress-strain curves (solid lines) and strain-hardening rate (SHR) curves (symbols) are shown in **Fig. 3B**. The uniform elongation (UEL, $\varepsilon_u$), as determined from the Considère criterion, is also shown for each alloy. As expected, the UTS of LPBF Ti-O significantly increases with oxygen content, from 347 MPa in Ti-0.09O to 915 MPa in Ti-0.45O. The nearly three-fold enhancement in UTS with only a 0.36wt.% increase in oxygen content highlights the potent hardening effect of oxygen interstitials in titanium, which is consistent with previous studies (*5, 6*). More importantly, despite a substantial increase in UTS, the UEL of LPBF Ti-O alloys is not adversely affected, in contrast to the strength-ductility trade-off phenomenon in conventional metallic materials (*25-27*). Indeed, the UEL of Ti-0.45O (15.6%) is 260% larger than that of Ti-0.09O (6.1%), which can be attributed to a much larger SHR in the specimen with higher oxygen content (**Fig. 3B**). In Ti-0.45O, the SHR even exhibits a three-stage evolution tendency, with a continuous upturn of SHR from true strain of $\varepsilon = 3.0\%$ to $\varepsilon = 7.0\%$ (**Fig. 3B**), delaying necking onset.

The simultaneous increase in strength and ductility with increasing oxygen content makes our LPBF Ti-0.33O and Ti-0.45O alloys stand out from the well-known 'banana' curves of metallic materials (*28, 29*). Compared with the literature data (*1, 2, 4, 9, 11-15, 30-37*) (**Fig. 3C**) for commercial purity (CP) titanium prepared by rolling/forging, and severe plastic deformation, as well as by LPBF, the LPBF Ti-0.33O and Ti-0.45O alloys in the present study show a superior synergy of strength and ductility. Notably, LPBF Ti-0.45O ($\sigma_y$=828 MPa and $\sigma_u$=895MPa, $\varepsilon_t$=29.7%) matches forged Ti-6Al-4V strength (ASTM B381 (*38*)) while tripling elongation (vs. 10%), offering a cost-efficient as-printed alternative.

**Deformation behavior of LPBF Ti-O alloys**

Multi-scale characterization of the deformation microstructure in Ti-0.09O and Ti-0.45O (**Fig. 4**) were conducted to reveal the strong-ductile synergy mechanism in LPBF Ti-O alloys. *Macro-scale* characterization shows homogeneous strain distribution in Ti-0.45O, even after a plastic strain of 10.0% (**Fig. 4C**), while the premature strain localization occurred in Ti-0.09O after a plastic strain of 4.0% (white arrows in **Fig. 4C**), explaining oxygen-enhanced strength-ductility synergy and delayed necking.

*Microscopically*, the initial microstructure of Ti-0.09O alloy was characterized by a high density of dislocations (**Fig. 4D**), typical for LPBF-processed titanium. Most of the dislocations were confirmed to be <*a*>-type according to a "g dot *b*" analysis under two-beam conditions (**Fig. 4, E and F**). Such a high density of dislocations, predominantly <*a*>-type, in the initial microstructure limits the scope for dislocation proliferation with further straining (**Fig. 4G**). In addition, most of the dislocations after deformation (at a plastic strain of 2.0%, **Fig. 4, H and I**) were also <*a*>-type. The lack of dislocation slip along the HCP <*c*> direction, together with the relatively high initial dislocation density results in a rapid decrease of the SHR and a low UEL in the Ti-0.09O alloy (**Fig. 4B**).

In contrast, only a low dislocation density was observed in the initial microstructure of Ti-0.45O (**Fig. 4J**), which can be attributed to recovery and/or recrystallization induced by the deliberately designed laser melting parameters (*19*). The initial low dislocation density endows the Ti-0.45O alloy with a continuous, extraordinary strain-hardening capability during tensile deformation (**Fig. 4K and L**). We observed a large number of dislocations, in a curved configuration, in the deformed microstructure of Ti-0.45O at a plastic strain of 6.0% (**Fig. 4L**). More importantly, most of these dislocations were confirmed to be <*c*+*a*>-type (**Fig. 4, M to P**), indicating extensive dislocation slip on pyramidal planes, in addition to <*a*>-type slip on prismatic planes. The promotion of <*c+a*>-type slip activity not only contributes to generalized plastic deformation, but also promotes dislocation interactions between those gliding on {11-22} pyramidal planes and {10-10} prismatic planes. Consequently, Ti-0.45O achieves a higher dislocation density (**Fig. 4**), corroborated by synchrotron X-ray diffraction revealing a rapid dislocation accumulation rate and order-of-magnitude higher density than Ti-0.09O at 14% strain (**fig. S11**). These results demonstrate oxygen-enhanced dislocation multiplication, which promotes superior strain hardening and homogeneous plastic strain distribution in high-oxygen Ti.

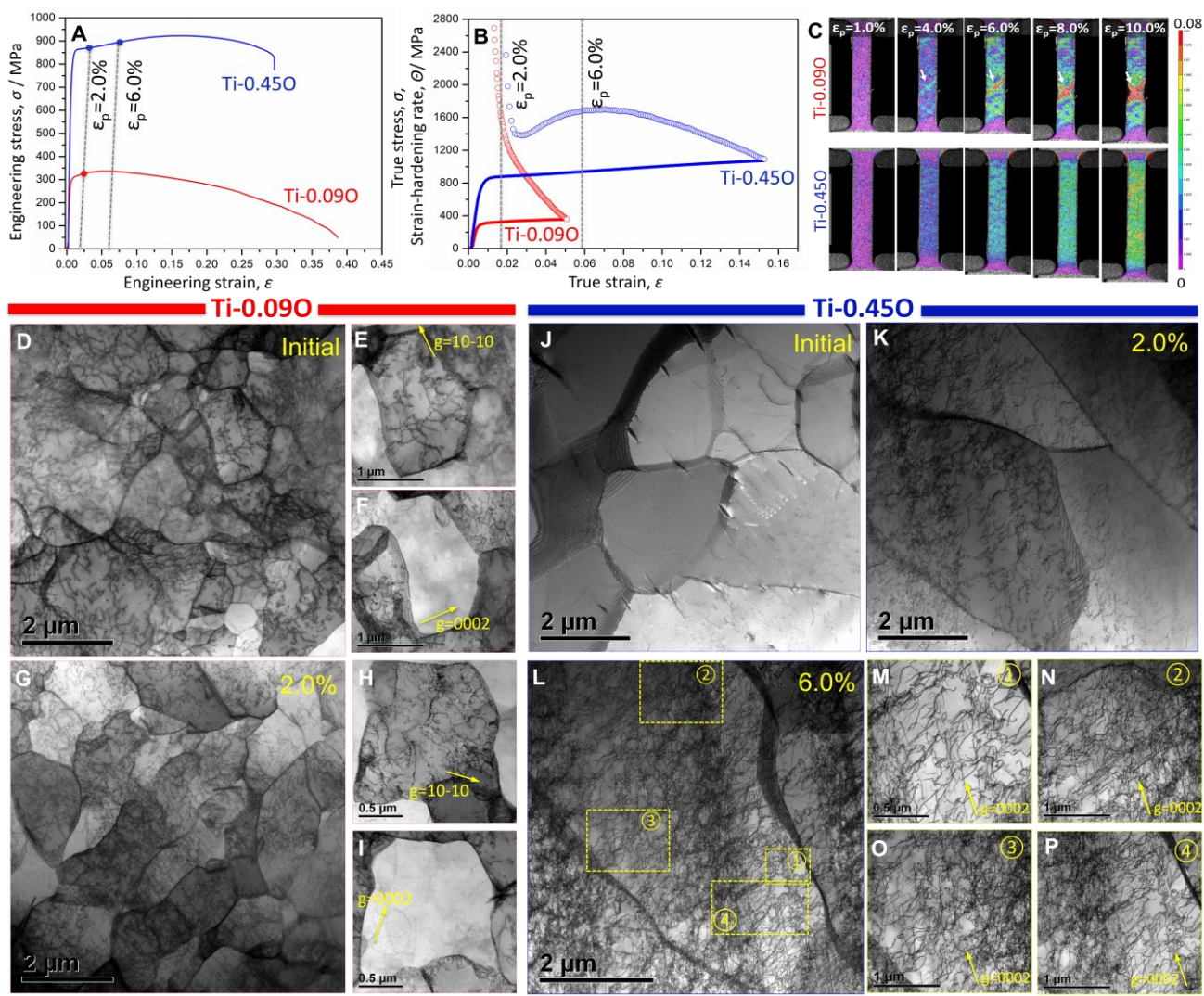

**Fig. 4. Deformation behavior of LPBF Ti-0.09O and Ti-0.45O**. (**A**) Engineering stress-strain curves of Ti-0.09O and Ti-0.45O, together with indications of the interrupted tensile strains. (**B**) True stress-strain and SHR curves of Ti-0.09O and Ti-0.45O. (**C**) Von-Mises strain distribution along the gauge part of both alloys up to a macro-tensile strain of 10%, (see scale bar on the right-hand side of the figure). Dislocation structures of Ti-

0.09O alloy in the initial state (**D** to **F**) and after a plastic strain of 2.0% (**G** to **I**). Bright-field (BF) images under a two-beam condition from selected areas using *g* = 10-10 (**E** and **H**) and *g* = 0002 (**F** and **I**). Dislocation structures of Ti-0.45O in the initial state (**J**) and after plastic strains of 2.0% (**K**) and 6.0% (**L** to **P**). BF images of the regions marked by ① - ④ in **L** collected under a two-beam condition using *g* = 0002 are shown in **M** to **P**, respectively.

**Origin of excellent strain-hardening ability in LPBF Ti-0.45O**

We systematically investigated the interaction processes between interstitial-type ZDSDs and two types of dislocations (<*a*>-type and <*c+a*>-type screw dislocations) in α-Ti, along with their corresponding stress-strain responses, to examine how ZDSDs interact with different slip systems. **Fig. 5A** illustrates the interaction between an interstitial-type ZDSD and a <*c+a*>-type screw dislocation; the interaction between an interstitial-type ZDSD and a <*a*>-type screw dislocation is shown in **fig. S12**

As a characteristic self-interstitial cluster, when a dislocation contacts a ZDSD, the ZDSD loses its intrinsic stacking domain and evolves into a dislocation line. These incorporated interstitials significantly impede further dislocation motion. For screw-type dislocations, after escape from the interstitial cluster, the original ZDSD transforms into a prismatic dislocation loop with the same Burgers vector as the screw dislocation, along with some isolated point defects. Notably, the resultant prismatic loop continues to impede the motion of subsequent <*c+a*>-type screw dislocations (**Fig. 5B**). This persistent obstruction occurs because both the ZDSD and prismatic loop fundamentally consist of self-interstitial clusters, despite their different initial configurations. Following interaction with the gliding dislocation, the structure of the prismatic loop is retained, which indicates a persistent and strong resistance of prismatic loops against subsequent dislocation slip.

The stress-strain curves in **Fig. 5C** characterize these interaction processes for both ZDSDs and prismatic loops. The structural evolution from initial ZDSD configuration (**Fig. 5D**) to post-interaction prismatic loop formation (**Fig. 5E**) suggests that substantial prismatic loop populations should be observable in after plastic deformation, as confirmed by experimental investigations (**Fig. 5F**).

The stress-strain curves reveal that ZDSDs significantly impede the mobility of both <*a*>-type and <*c+a*>-type screw dislocations. For <*a*>-type screw dislocations, the required shear stress to overcome the obstruction of ZDSD increases from 300 MPa (without ZDSD interaction) to a maximum of 1100 MPa (with ZDSD interaction). Similarly, <c+a>-type screw dislocations exhibit an increased stress requirement from 500 MPa to 850 MPa when interacting with ZDSDs. These quantitative comparisons demonstrate that ZDSDs substantially increase the resistance to dislocation motion for all slip systems, with a more pronounced effect on <a>-type than <c+a>-type screw dislocations. The comparable resistance between <a>-type and <c+a>-type screw dislocations in the presence of ZDSD explains why a higher fraction of <c+a>-type dislocations are present after deformationLPBF Ti-0.45O compared to LPBF Ti-0.09O.

The limited plasticity of Ti primarily stems from the significant disparity in activation capabilities between different slip systems; <a>-type dislocations exhibit high mobility while <c+a>-type dislocations show considerably lower mobility, leading to heterogeneous plastic deformation along different orientations. The presence of ZDSDs not only increases the resistance required for dislocation glide but also helps balance the mobility difference between various slip systems, Therby contributing to the simultaneous improvement of both strength and ductility.

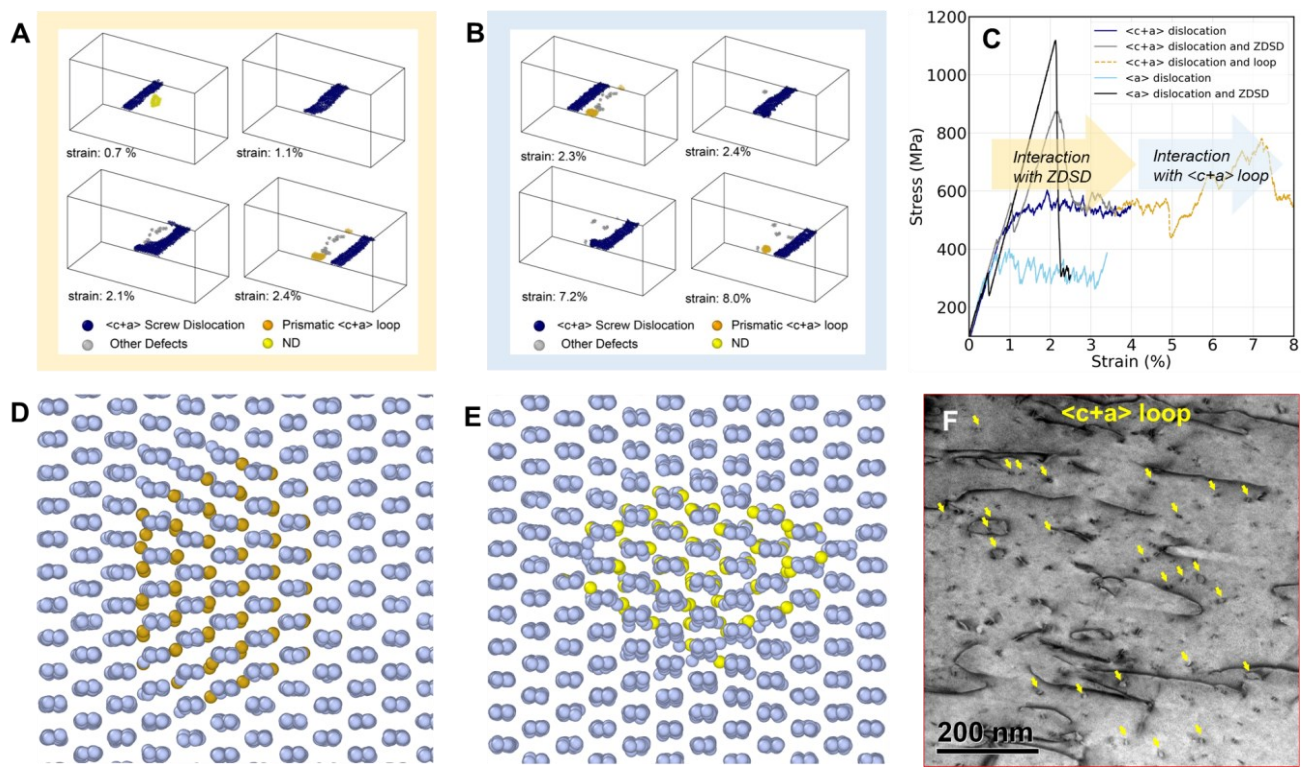

**Fig. 5. Interaction between dislocations and ZDSDs.** Molecular dynamics (MD) simulations of **(A)** interaction between a <c+a>-type dislocation and an ordered oxygen ZDSD; **(B)** interaction between a <c+a>-type dislocation and a prismatic **<c+a>** loop; **(C)** Stress-strain curves of the interactions between <c+a>-type

dislocations and ZDSD and prismatic <*c+a*> loops, as well as between <a>-type dislocations and ZDSDs. **(D)** Atomic configuration of a ZDSD in the MD simulations. **(E)** Atomic configuration of aprismatic <*c+a*> loop after the passage of a <*c+a*>-type dislocation in the MD simulation. **(F)** Typical dislocation structure in LPBF Ti-0.45O tensile deformed by a plastic strain of 2%. A large number of prismatic <*c+a*> loops (indicated by yellow arrows) can be observed.

## Conclusions

In summary, by control of LPBF processing parameters, we achieved an exceptional combination of ultimate tensile strength (UTS, $\sigma_u$=915 MPa) and uniform elongation (UEL, $\varepsilon_u$ =15.6%) in a high-oxygen titanium (Ti-0.45O) at room temperature. Moreover, both the UTS and UEL of LPBF Ti-O alloys increased with increasing oxygen content, evading the traditional trade-off between strength and ductility of structural materials. Through in-depth microstructural characterization and computational simulations, the excellent balance between strength and ductility in LPBF Ti-0.45O alloy is attributed to a new class of crystal defect, namely zero-dimensional stacking domains, present in a high density in the LPBF Ti-O alloys. These coherent stacking domains, stabilized by oxygen interstitials, contribute to the strain hardening of LPBF Ti-O alloys in two both by promoting <*c+a*>-type dislocation slip, and also by providing barriers to continued <*c+a*>-type slip during further plastic deformation.

The LPBF Ti-O alloys in this study, with both high strength and large ductility, offer an alternative solution to conventional biomedical titanium alloys that rely on expensive alloying elements such as vanadium or niobium, as well as toxic alloying elements such as aluminum. In addition, net-shaped products with irregular geometries can also be directly realized using LPBF. The results also highlight the potential for in-situ oxygen alloying without sacrificing the ductility, thereby opening up new possibilities for interstitial solute engineering in a wide range of metal materials by additive manufacturing.

**Acknowledgments**

The authors would like to acknowledge the financial support from National Magnetic Confinement Fusion Energy R&D Program of China [Grant number 2019YFE03130003, 2022YEF03130003]; National Natural Science Foundation of China [Grant number 52105373, 52275391, 52304403]; and China Postdoctoral Science Foundation [Grant number 2022M711753]. Y.C. acknowledges the support of JSPS KAKENHI (No. JP24K08018). The synchrotron X-ray diffraction experiments were performed at BL13XU in SPring-8, Harima, Japan under the beam time number of 2023B1762. Additionally, we thank Mingshen Li (School of Materials Science and Engineering, Tsinghua University) for manuscript feedback, and Yuxiang Wang (Analytical and Testing Center, Northeastern University) for tensile testing.

**Competing interests:** The authors declare that they have no competing financial interests.

**Data and materials availability:** Data is available via request to the corresponding author.

manufacturing of Ti-6Al-4V with forging standard out-of-plane tensile properties. *Journal of Materials Processing Technology* 322, 118201 (2023).